\documentclass[aip, jcp, reprint]{revtex4-2}
\usepackage{graphicx}
\usepackage{multirow}
\usepackage{amsmath, bm}
\usepackage{amssymb}
\usepackage{mathtools}
\usepackage{xcolor}
\usepackage[normalem]{ulem}
\usepackage[version=4]{mhchem}
\usepackage[ruled,linesnumbered,vlined]{algorithm2e}
\usepackage{subcaption}
\SetAlgoCaptionSeparator{.}
\SetAlCapHSkip{0pt}
\DontPrintSemicolon
\SetStartEndCondition{ }{}{}%
\SetKwProg{Fn}{def}{\string:}{}
\SetKwProg{CUDAFn}{def __global__}{\string:}{}
\SetKwFunction{Range}{range}
\SetKw{KwTo}{in}
\SetKwFor{For}{for}{\string:}{}%
\SetKwFor{In}{in CUDA block}{\string:}{}%
\SetKwIF{If}{ElseIf}{Else}{if}{:}{elif}{else:}{}%
\SetKwFor{While}{while}{:}{fintq}%
\SetKwFunction{Append}{.append}
\SetKwFunction{Enumerate}{enumerate}
\SetKwFunction{JKkernel}{jk\_kernel}
\SetKwFunction{Rysroot}{rys\_roots<$N$>}
\SetKwFunction{RR}{rr<$N$>}
\SetKwFunction{RRip}{rr1<$N$>}
\SetKwFunction{AtomicAdd}{atomicAdd}
\SetKwComment{Comment}{\#}{}
\newcommand{\var}{\texttt}

\makeatletter
\newcommand{\AlgoCaptionFormat}{}
\newcommand{\SetAlgoCaptionFormat}[1]{\def\AlgoCaptionFormat{#1}}
\renewcommand{\algocf@makecaption@ruled}[2]{%
    \global\sbox\algocf@capbox{\hskip\AlCapHSkip%
        \setlength{\hsize}{\columnwidth}
        \addtolength{\hsize}{-2\AlCapHSkip}
        \vtop{\AlgoCaptionFormat\algocf@captiontext{#1}{#2}}}
}%
\makeatother

\SetAlgoCaptionFormat{\raggedright}
\usepackage{comment}

\newcommand{\pyscf}{\textsc{PySCF}\xspace}
\newcommand{\gpupyscf}{\textsc{GPU4PySCF}\xspace}

\begin{document}

\title{Implementation of the multigrid Gaussian-Plane-Wave algorithm with GPU acceleration in \textsc{PySCF}}

\author{Rui Li}
\affiliation{Division of Chemistry and Chemical Engineering,
	California Institute of Technology, Pasadena, CA 91125, United States of America}
\affiliation{Marcus Center for Theoretical Chemistry,
	California Institute of Technology, Pasadena, CA 91125, United States of America}

\author{Xing Zhang}
\affiliation{Division of Chemistry and Chemical Engineering,
	California Institute of Technology, Pasadena, CA 91125, United States of America}
\affiliation{Marcus Center for Theoretical Chemistry,
	California Institute of Technology, Pasadena, CA 91125, United States of America}

\author{Qiming Sun}
\affiliation{Bytedance Seed, Bellevue, WA 98004, United States of America}

\author{Yuanheng Wang}
\affiliation{Bytedance Seed, Beijing, China} 

\author{Junjie Yang}
\affiliation{Division of Chemistry and Chemical Engineering,
	California Institute of Technology, Pasadena, CA 91125, United States of America}
\affiliation{Marcus Center for Theoretical Chemistry,
	California Institute of Technology, Pasadena, CA 91125, United States of America}

\author{Garnet Kin-Lic Chan$^*$}
\affiliation{Division of Chemistry and Chemical Engineering,
	California Institute of Technology, Pasadena, CA 91125, United States of America}
\affiliation{Marcus Center for Theoretical Chemistry,
	California Institute of Technology, Pasadena, CA 91125, United States of America}

\begin{abstract}
	\textbf{ABSTRACT:}
	We introduce a GPU-accelerated multigrid Gaussian-Plane-Wave density fitting (FFTDF) approach for efficient Fock builds
	and nuclear gradient evaluations within Kohn–Sham density functional theory, as implemented in the \gpupyscf module of \pyscf.
	Our CUDA kernels employ a grid-based parallelization strategy for contracting Gaussian basis function pairs
	and achieve up to $80$\% of the FP64 peak performance on NVIDIA GPUs, with no loss of efficiency for high angular momentum (up to $f$-shell) functions.
	Benchmark calculations on molecules and solids with up to 1536 atoms and 20480 basis functions show up to $25\times$ speedup on an H100 GPU relative to the CPU implementation on a 28-core shared memory node.
	For a 256-water cluster, the ground-state energy and nuclear gradients can be computed in $\sim$30 seconds 
	on a single H100 GPU.
	This implementation serves as an open-source foundation for many applications, such as \emph{ab initio} molecular dynamics and high-throughput calculations.
	\label{abstract}
\end{abstract}

\maketitle

\section{Introduction}
Graphics processing units (GPUs) have enabled orders-of-magnitude speedups for numerous computational chemistry methods,
including molecular dynamics,\cite{sakamaki412ImplementationEvaluation2009, lee2018gpu, fanGPUMDPackageConstructing2022, eastman2023openmm} Kohn-Sham density functional theory (KS-DFT),\cite{yasuda2008accelerating, ufimtsevGraphicalProcessingUnits2008, ufimtsevQuantumChemistryGraphical2008, ufimtsevQuantumChemistryGraphical2009, ufimtsevQuantumChemistryGraphical2009a, luehrDynamicPrecisionElectron2011, haceneAcceleratingVASPElectronic2012, miaoAccelerationElectronRepulsion2013, maintzCuVASPGPUAcceleratedPlaneWave2012, kussmannHybridCPUGPU2017, romeroPerformanceStudyQuantum2018, seritanTeraChemGraphicalProcessing2021, barcaFasterSelfConsistentField2021, qiHybridCPUGPU2023, wangExtendingGPUacceleratedGaussian2024, stocks2025efficient, wu2025enhancing, ambrogioAcceleratedLinearAlgebra2025} correlation theories such as second-order M{\o}ller-Plesset perturbation theory,\cite{vogt2008accelerating, terachemMP2, stocksHighPerformanceMultiGPUAnalytic2024} coupled-cluster theory,\cite{deprince2011coupled, snyderDirectcompatibleFormulationCoupled2017, datta2023accelerating, hohensteinGPUAccelerationRankreduced2021} and many others.\cite{yanGraphicsProcessingUnit2013, afqmc, afqmc2, dmrg}
This is because of the distinct design of GPU chips, which compared to central processing units (CPUs), offer significantly higher instruction throughput and memory bandwidth.\cite{cuda}

Nevertheless, fully exploiting GPU throughput generally requires substantial algorithmic redesign to accommodate the distinct aspects of GPU architectures. GPU kernel performance can deteriorate significantly when implementations do not carefully follow best practices, such as minimizing register spilling and global-memory traffic.\cite{cuda}
For example, in our GPU implementation of two-electron repulsion integrals,\cite{liIntroducingGPUAcceleration2025} we found a strong dependence of the kernel efficiency on the angular momentum of the Gaussian basis functions, because higher angular momenta require deeper recursion relations, leading to intermediates that exceed the available register capacity.
Addressing such challenges is the key to achieving efficient GPU implementations of quantum chemistry methods.

In this work, we focus on accelerating pseudopotential KS-DFT on GPUs by employing the multigrid Gaussian-Plane-Wave density fitting (denoted FFTDF in \pyscf\cite{pyscf}) approach, originally introduced by Lippert et al.\cite{lippert1997hybrid,lippert1999gaussian}
Compared with the original CPU implementation in \pyscf (which has already enabled several large-scale DFT applications\cite{white2023quantum,smyser2024use,li2025image}),
our GPU implementation introduces an additional layer of parallelism at the real-space grid level.
In particular, the uniform grid in the FFTDF method maps naturally onto the hierarchical thread-block structure of GPUs, allowing for a balanced workload distribution and to minimize global memory traffic so that the kernels remain compute-bound. Our design achieves near-peak FP64 throughput on NVIDIA GPUs and offers state-of-the-art performance for KS-DFT Fock builds with local and semi-local exchange-correlation (XC) functionals. 
Overall, our implementation supports the local density approximation (LDA), generalized gradient approximation (GGA), and meta-generalized gradient approximation (meta-GGA) functionals. Both $\Gamma$-point and $k$-point sampling are available, along with band structure analysis for orthorhombic and non-orthorhombic periodic systems.

The remainder of the paper is organized as follows.
Section \ref{sec:background} provides a brief overview of the multigrid FFTDF algorithm.
Section \ref{sec:implementation} describes our GPU implementation in detail.
Section \ref{sec:result} reports timing results and kernel-level performance analyses for a range of molecules and solids. Finally, Section \ref{sec:conclusion} summarizes our conclusions.

\section{Background \label{sec:background}}

In this section, we review the multigrid FFTDF approach for KS-DFT Fock builds.
In FFTDF, the underlying single particle basis functions are chosen as  crystalline Gaussian type orbitals (GTOs),
\begin{equation}
	\phi_{\mu,\mathbf{k}} (\mathbf{r}) = \sum_{\mathbf{T}} e^ {i\mathbf{k} \cdot \mathbf{r} } \phi^\mathbf{T}_\mu(\mathbf{r}),
\end{equation}
where $\mathbf{T}$ is a lattice translation vector,
and $\mathbf{k}$ is a crystal momentum vector in the first Brillouin zone. $\phi_\mu(\mathbf{r})$ represents the Gaussian basis functions located in the reference unit cell and
\begin{equation}
	\phi^{\mathbf{T}}_\mu(\mathbf{r}) = \phi_\mu (\mathbf{r - T}).
\end{equation}
In this paper, we only describe $\Gamma$-point calculations, i.e., $\mathbf{k} = \mathbf{0}$, for simplicity, although our implementation in the \gpupyscf module fully supports $k$-point sampling.

The real-space electron density can be expressed as
\begin{equation}\label{eq:rho}
	\rho(\mathbf{r}) = \sum_{\mathbf{S}\mathbf{T}} \sum_{\mu \nu} D_{\mu\nu}
	\phi^{\mathbf{S}}_\mu (\mathbf{r}) \phi^{\mathbf{T}}_\nu (\mathbf{r}) \;,
\end{equation}
where $\mathbf{D}$ is the one-electron density matrix.
The two-electron part of the Fock matrix, including the Coulomb and XC contributions, is then evaluated numerically in real space as
\begin{equation} \label{eq:J}
	F^\text{2e}_{\mu\nu} = \sum_{\mathbf{S}\mathbf{T}} \int_\Omega \phi^{\mathbf{S}}_\mu (\mathbf{r}) v_{\text{Hxc}} (\mathbf{r}) \phi_\nu^{\mathbf{T}} (\mathbf{r}) d\mathbf{r}\;,
\end{equation}
where
\begin{equation}\label{eq:vH}
	v_{\text{Hxc}} (\mathbf{r}) = \mathcal{F}^{-1}\big(\tilde{v}_{\text{Hxc}} (\mathbf{G})\big) =
	\frac{1}{\Omega} \sum_{\mathbf{G} \neq \mathbf{0}} \tilde{v}_{\text{Hxc}} (\mathbf{G}) e^{i\mathbf{G}\cdot\mathbf{r}}
\end{equation}
and
\begin{equation} \label{eq:vHG}
	\tilde{v}_{\text{Hxc}} (\mathbf{G}) = \frac{4\pi}{|\mathbf{G}|^2} \tilde{\rho}(\mathbf{G}) + \tilde{v}_\text{xc}
\end{equation}
are the Hartree exchange-correlation (Hxc) potential and its Fourier representation, with $\tilde{\rho}$ the Fourier representation of $\rho$,
$\tilde{v}_\text{xc}$ the Fourier representation of the XC potential
, $\mathcal{F}^{-1}$ the inverse of Fourier transformation $\mathcal{F}$,
$\mathbf{G}$ the momentum vectors of the plane waves,
and $\Omega$ the unit cell volume.
The first term in Eq.~\eqref{eq:vHG} is the Hartree potential in Fourier space.
Note that the Fourier transforms are performed using the fast Fourier transform (FFT), while the plane waves can be viewed as a density fitting basis, hence the designation FFTDF.

In conventional Gaussian basis sets (including the ones used in this work)
the underlying GTO exponents typically span a wide range, yielding both compact and diffuse orbitals.
It is therefore beneficial to employ multiple sets of numerical grids with varying resolutions when evaluating Eqs.~\eqref{eq:rho}--\eqref{eq:vH}.
This leads to the multigrid method, where the GTO pair products
(which are also GTOs) are sorted according to
their exponents and subsequently binned into groups.
Each group is assigned a distinct uniform grid, with an associated plane wave cutoff $G_{\alpha}$.\cite{vandevondele2005quickstep}
On each grid, Eq.~\eqref{eq:rho} is evaluated for the corresponding group of GTO pairs to obtain the partial density $\rho_{\alpha}$.
The total electron density is then accumulated (on the finest grid) by Fourier interpolation as
\begin{equation}
	\tilde{\rho}(\mathbf{G}) = \sum_{\alpha} \tilde{\rho}_{\alpha}(\mathbf{G})\;,
\end{equation}
where
\begin{equation}
	\tilde{\rho}_{\alpha}(\mathbf{G}) =
	\begin{cases}
		\mathcal{F} \big(\rho_{\alpha}(\mathbf{r}) \big) & \text{if }|\mathbf{G}| < G_{\alpha} \;,    \\
		0                                                & \text{if } |\mathbf{G}| \ge G_{\alpha} \;.
	\end{cases}
\end{equation}
Once $\tilde{\rho}$ is obtained, the Hxc potential is computed in Fourier space [Eq.~\eqref{eq:vHG}],
transformed back to real space [Eq.~\eqref{eq:vH}]
for each of the multiple grids, and integrated numerically via Eq.~\eqref{eq:J} to yield its contribution to the Fock matrix.

For LDA functionals, the XC potential $v_\text{xc}^{\text{LDA}}[\rho]$ depends only on the electron density $\rho$, and
\begin{equation}
	v_{\text{xc}}^{\text{LDA}}(\mathbf{r}) = \frac{\delta E_{\text{xc}}^{\text{LDA}}}{\delta \rho(\mathbf{r})} \;,
\end{equation}
where $E_{\text{xc}}^{\text{LDA}}$ is the corresponding XC energy, and $\delta$ denotes the functional derivative.

For GGA functionals, the XC potential $v_\text{xc}^{\text{GGA}}[\rho, \bm{\nabla} \rho]$ is a functional of both the electron density and its gradient.
Due to the cost of computing the density gradient in real space (which requires basis function gradients), it is approximated by a gradient evaluation in Fourier space followed by an inverse Fourier transform,
\begin{equation}
	\bm{\nabla} \rho(\mathbf{r}) = \mathcal{F}^{-1} \big(i \mathbf{G} \tilde{\rho} (\mathbf{G}) \big) \;.
\end{equation}

In addition to the LDA contribution, in the GGA formalism, evaluating the Fock matrix involves an integration over the XC potential of the form
\begin{equation}\label{eq:vxc_GGA}
V_{\mu\nu}^{\text{xc,GGA}} = \sum_{\mathbf{S}\mathbf{T}} \int_\Omega \bm{\nabla} \big(\phi^\mathbf{S}_\mu (\mathbf{r}) \phi_\nu^{\mathbf{T}} (\mathbf{r}) \big) \cdot \bm{v}_{\text{xc}}^{\text{GGA}}(\mathbf{r}) \mathbf{dr},
\end{equation}
where
\begin{equation}
\bm{v}_{\text{xc}}^{\text{GGA}}(\mathbf{r}) = \frac{\delta E_\text{xc}^\text{GGA}}{\delta \bm{\nabla} \rho(\mathbf{r})}.
\end{equation}
In conventional XC matrix evaluation programs based on general (non-uniform) numerical integration grids, evaluating this integral requires computing gradients of each orbital pair $\phi_\mu^{\mathbf{S}} (\mathbf{r}) \phi_\nu^{\mathbf{T}} (\mathbf{r})$, which is computationally demanding.
Applying integration by parts to Eq.~\eqref{eq:vxc_GGA} gives
\begin{equation}
V_{\mu\nu}^{\text{xc,GGA}} = \sum_{\mathbf{S}\mathbf{T}} \int_\Omega -\phi^{\mathbf{S}}_\mu (\mathbf{r}) \phi_\nu^{\mathbf{T}} (\mathbf{r}) \bm{\nabla} \cdot \bm{v}_{\text{xc}}^{\text{GGA}}(\mathbf{r})\mathbf{dr},
\end{equation}
where gradients of the real-space XC potential can be efficiently approximated using Fourier transforms, in the same manner as those of the electron density,
\begin{equation}\label{eq:vxc_GGA_1}
\bm{\nabla} \cdot \bm{v}_\text{xc}^{\text{GGA}}(\mathbf{r}) = \mathcal{F}^{-1}\left( i\mathbf{G} \cdot \mathcal{F}(\bm{v}_{\text{xc}}^{\text{GGA}}(\mathbf{r})) \right).
\end{equation}
In practice, Eq.~\eqref{eq:vxc_GGA_1} is absorbed into the real-space Hxc potential in Eq.~\eqref{eq:vH}, making computation of the Fock matrix for GGA functionals as inexpensive as for LDA functionals.

However, such approximations cannot be directly applied to meta-GGA functionals, due to their dependence on
the kinetic energy density
\begin{equation}
    \tau(\mathbf{r}) = \frac{1}{2} \sum_{\mathbf{S}\mathbf{T}} \sum_{\mu \nu} D_{\mu\nu}
	\bm{\nabla} \phi^\mathbf{S}_\mu (\mathbf{r}) \cdot 	\bm{\nabla} \phi^{\mathbf{T}}_\nu (\mathbf{r}) \;.
\end{equation}
The corresponding Fock matrix contains an additional term that requires explicit evaluation of orbital gradients:
\begin{equation} \label{eq:J_metaGGA}
\begin{aligned}
	F^\text{2e}_{\mu\nu} = \sum_{\mathbf{S}\mathbf{T}} \int_\Omega \mathbf{dr} \bigg( &\phi^\mathbf{S}_\mu (\mathbf{r}) v_{\text{Hxc}} (\mathbf{r}) \phi_\nu^{\mathbf{T}} (\mathbf{r}) \\
    &\quad + v_{\text{xc}}^{\text{mGGA}} (\mathbf{r}) \bm{\nabla} \phi^\mathbf{S}_\mu (\mathbf{r})  \cdot \bm{\nabla}\phi_\nu^{\mathbf{T}} (\mathbf{r})\bigg) \;,
    \end{aligned}
\end{equation}
where
\begin{equation}
    v_{\text{xc}}^{\text{mGGA}} (\mathbf{r}) = \frac{\delta E_\text{xc}^\text{mGGA}}{\delta \tau(\mathbf{r})} \;.
\end{equation}
Eq.~\eqref{eq:J_metaGGA} replaces Eq.~\eqref{eq:J} when a meta-GGA functional is required.
The XC energy and potential are conveniently computed using the \textsc{LibXC}\cite{lehtola2018recent} library.

Finally, the remaining one-electron contributions, including the kinetic energy and the GTH pseudopotential,\cite{hartwigsen1998relativistic} are precomputed and cached at the beginning of the self-consistent field iterations.
The kinetic energy, as well as the short-range local and non-local components of the GTH pseudopotential are evaluated analytically.
The long-range part of the GTH pseudopotential is treated as arising from a Gaussian charge distribution, as described in Eq.~10 of Ref.~\citenum{vandevondele2005quickstep}, and is incorporated into the Hxc potential in Eq.~\eqref{eq:vHG} for the evaluation of the Fock matrix.

\section{Implementation \label{sec:implementation}}
In this section, we present our GPU implementation of the multigrid FFTDF approach, preceded by a brief review of the CPU implementation to clarify key algorithmic distinctions. 

\subsection{CPU}
The Fock build consists of two stages, namely, construction of the electron density [Eq.~\eqref{eq:rho}] and integration of the Hxc potential [Eq.~\eqref{eq:J}], both performed in real space.
The central computational task in both cases is the evaluation of GTO pair products on uniform grids.
For clarity, we omit the explicit notation of lattice translations in the following discussion. 
However, in the CPU implementation, we adopt the convention that one of the orbitals in Eqs.~\eqref{eq:rho} and \eqref{eq:J} is fixed in the reference unit cell, while the other is summed over its periodic images. Contributions arising outside the reference cell are subsequently folded back into it. 
In other words, Eqs.~\eqref{eq:rho} and \eqref{eq:J} are now computed as
\begin{equation} \label{eq:rho_wrap}
    \rho(\mathbf{r}) = \sum_\mathbf{S} \Big( 
    \sum_\mathbf{T} D_{\mu\nu} \phi^\mathbf{S}_\mu(\mathbf{r}) \phi^\mathbf{S+T}_\nu(\mathbf{r})
    \Big) \;,
\end{equation}
and
\begin{equation} \label{eq:J_wrap}
    F^\text{2e}_{\mu\nu} = \sum_\mathbf{S} \Big(
    \sum_\mathbf{T} \int_\Omega \phi^\mathbf{S}_\mu(\mathbf{r}) v_\text{Hxc}(\mathbf{r}) \phi^\mathbf{S+T}_\nu(\mathbf{r}) d\mathbf{r}
    \Big) \;.
\end{equation}
This choice simplifies the screening procedure and reduces the computational cost.

First, to achieve a linear scaling algorithm, proper prescreening of GTO pairs is necessary.
For each shell of GTOs, we compute a shell cutoff radius $r_\text{cut}$, such that
\begin{equation} \label{eq:cut}
	r_\text{cut}^2\phi_\text{R}(r_\text{cut}) < \tau \;,
\end{equation}
where $\phi_\text{R}$ is the radial part of a primitive Gaussian function, and $\tau$ is a predefined accuracy threshold.
Only those GTOs that overlap with each other (i.e., the distance between them is smaller than the sum of their shell cutoff radii) are considered when evaluating Eqs.~\eqref{eq:rho} and \eqref{eq:J}.

Next, we follow the strategies of VandeVondele et al.\cite{vandevondele2005quickstep} to compute GTO pair products on uniform grids.
It takes $O(N_\text{pair}N_\text{grid})$ floating-point operations (FLOPs), where $N_\text{pair}$ and $N_\text{grid}$ are the numbers of GTO pairs and grid points, respectively.
Note that $N_\text{pair}$ grows only linearly with system size due to the prescreening
mentioned above, while $N_\text{grid}$ remains constant as we further eliminate non-contributing grid points for each GTO pair (see below).

In orthorhombic lattices, primitive Cartesian Gaussian basis functions
(and their products) evaluated on uniform grids admit simple separable factorizations into $x$-, $y$-, and $z$-dependent components,
each of which can be computed independently.
(An analogous factorization applies to non-orthorhombic lattices, documented in the Appendix, Eq.\ref{eq:exp_rr1_non_orthogonal_r1}-\ref{eq:exp_rr1_non_orthogonal_r3}.)
For example, a single primitive Gaussian factorizes as
\begin{equation}
	g_a(\mathbf{r}) \equiv g_a(x,y,z) = g_a(x)g_a(y)g_a(z) \;,
\end{equation}
where
\begin{equation} \label{eq:gx}
	g_a(x) = (x-A_x)^{l_x^a} e^{-\alpha_a (x-A_x)^2} \;,\ldots
\end{equation}
For the product of a Gaussian pair, Eq.~\eqref{eq:gx} becomes
\begin{equation} \label{eq:gab}
	g_{ab}^{l_x^a l_x^b}(x) = (x-A_x)^{l_x^a} (x-B_x)^{l_x^b} e^{-\alpha_p (x-P_x)^2} \;,
\end{equation}
where $\alpha_p = \alpha_a+\alpha_b$ and $\mathbf{P} = (\alpha_a \mathbf{A} + \alpha_b \mathbf{B}) / \alpha_p$. In Eq.~\eqref{eq:gab}, the grid-point independent prefactor is omitted for clarity.
Such factorization offers the benefit that at most $O(\sqrt[3]{N_\text{grid}})$ evaluations of Gaussian functions [Eq.~\eqref{eq:gab}] are needed, compared with
$O(N_\text{grid})$ evaluations for the non-factorized case.
Furthermore, the uniform grid spacing allows one to use the following recursion relation to reduce the number of exponential function (\texttt{exp}) evaluations to only three per dimension:
\begin{align} \label{eq:exp_rr}
	g_{ab}^{00}(x_{i+1}) & = g_{ab}^{00}(x_i) e^{-(2i+1)\alpha_p dx^2} e^{-2\alpha_p dx {(x_0 - P_x)}}\;, \\ \label{eq:exp_rr1}
	g_{ab}^{00}(x_0)     & = e^{-\alpha_p {(x_0 - P_x)}^2} \;,
\end{align}
where $x_0$ denotes the grid point nearest to the Gaussian center $P_x$, $dx$ is the grid spacing, and $x_i = x_0+i\times dx$.
In Eq.~\eqref{eq:exp_rr}, $i$ (which can be negative) iterates over the grid points within a sphere centered
at $\mathbf{P}$, whose radius is determined through Eq.~\eqref{eq:cut} for each Gaussian pair.
Because evaluating \texttt{exp} is one to two orders of magnitude more
expensive than a multiplication, the procedure above substantially reduces the overall computational cost.
In addition, the single-lattice-sum-plus-folding scheme [Eqs.~\eqref{eq:rho_wrap} and \eqref{eq:J_wrap}] requires significantly fewer Gaussian pair (more precisely, \texttt{exp}) evaluations than
the double-lattice-sum formalism [Eqs.~\eqref{eq:rho} and \eqref{eq:J}], which justifies our adoption of the former approach.

Finally, for orbital pairs within the same shell pair,
common factors in the polynomial prefactor of Eq.~\eqref{eq:gab} can be extracted via a binomial expansion,
\begin{align} \label{eq:binomial}
	 & (x-A_x)^{l_x^a} (x-B_x)^{l_x^b} \nonumber \\
	 & = \sum_{k=0}^{l_x^b}
	\left(\begin{array}{l} l_x^b \\ k \end{array} \right)
	(x-A_x)^{k+l_x^a} (A_x-B_x)^{l_x^b-k} \;,
\end{align}
where $l_x^a \in \{0, 1, \ldots, l^a\}$, and
$l_x^b \in \{0, 1, \ldots, l^b\}$ for shells $a$ and $b$ with angular momenta
$l^a$ and $l^b$, respectively.
Consequently, instead of evaluating Eq.~\eqref{eq:gab} for every $(l_x^a, l_x^b)$ pair, i.e., $l^al^b$ times
per grid point along a given dimension, only $(l^a+l^b+1)$ distinct functions need to be computed, namely,
\begin{equation} \label{eq:gab_lp}
	g_{ab}^{l_x^p}(x) = (x-A_x)^{l_x^p} e^{-\alpha_p (x-P_x)^2} \;,
\end{equation}
where $l_x^p \in \{0, 1, \ldots, l^a+l^b\}$.
The resulting FLOP reduction is more evident in the contraction in Eq.~\eqref{eq:rho}, where the contribution from each shell pair can now be written as
\begin{equation} \label{eq:rho1}
	  \sum_{l_x^p,l_y^p,l_z^p=0}^{l^a+l^b}
	\mathcal{D}_{l_x^p,l_y^p, l_z^p} g_{ab}^{l_x^p} (x) g_{ab}^{l_y^p} (y) g_{ab}^{l_z^p} (z) \;,
\end{equation}
and $\mathcal{D}$ involves contractions between the density matrix $D$ in Eq.~\eqref{eq:rho} and the expansion coefficients in
Eq.~\eqref{eq:binomial}.
The computational cost of evaluating Eq.~\eqref{eq:rho1} therefore scales as $O((l^a+l^b)N_\text{grid})$,
which is orders of magnitude lower than the
$O((l^a l^b)^2 N_\text{grid})$
scaling of the original expression in Eq.~\eqref{eq:rho} for shells with higher angular momenta.

The same conclusion applies to the integration of the Hxc potential in Eq.~\eqref{eq:J}.
In this case, an intermediate 3-dimensional tensor $\mathcal{V}_{l_x^p,l_y^p, l_z^p}$ 
[analogous to $\mathcal{D}$ in Eq.~\eqref{eq:rho1}, with
$0 \leq l_x^p, l_y^p, l_z^p \leq l^a+l^b$] is first constructed for each shell pair by contracting the real-space potential with the Gaussian products defined in Eq.~\eqref{eq:gab_lp}:
\begin{equation}
    \mathcal{V}_{l_x^p,l_y^p, l_z^p} = 
    \frac{\Omega}{N_\text{grid}}\sum_{x, y, z} v_\text{Hxc}(x, y, z)
	 g_{ab}^{l_x^p} (x) g_{ab}^{l_y^p} (y) g_{ab}^{l_z^p} (z) \;,
\end{equation}
The computational cost of this step scales as $O((l^a+l^b)N_\text{grid})$.
The obtained tensor $\mathcal{V}$ is then contracted with the expansion coefficients in Eq.~\eqref{eq:binomial} along each Cartesian component to assemble the final Fock matrix.

Our CPU implementation of the multigrid FFTDF approach employs straightforward OpenMP parallelization over Gaussian shell pairs,
without explicit load-balancing optimization.
We also note that Eq.~\eqref{eq:rho1} leads to skinny matrix multiplications for basis functions with low angular momentum,
a regime in which standard BLAS level-3 routines
(used exclusively in our current implementation) might be suboptimal.
As such, additional performance gains are likely possible through manually unrolled contraction loops combined with cache-aware blocking and explicit SIMD (single instruction, multiple data) vectorization.
Later, we provide a performance comparison of our implementation with the leading implementation of multigrid KS-DFT (which performs such optimizations) in CP2K.\cite{Schutt2016-sx,Kuhne2020-tt,Heinecke2016-md,Hutter2014-tx,Borstnik2014-kx,Marek2014-gh,vandevondele2005quickstep,Krack2005-wx,Frigo2005-cq,hartwigsen1998relativistic,lippert1997hybrid,Goedecker1996-le}

\subsection{GPU}

In the CPU implementation, the intermediate Gaussian products in Eq.~\eqref{eq:gab_lp} are precomputed and cached in memory.
In our initial GPU implementation, the same strategy was adopted, with these intermediates evaluated and stored in global memory.
However, this approach results in substantial global memory traffic due to repeated reads of the intermediates, as well as excessive write conflicts during reductions over orbital indices or grid points.
As a result, the GPU implementation achieved only marginal speedup compared to the CPU version.

In various GPU optimization practices, minimizing global-memory access is often more critical for performance than minimizing the FLOP count.
In contrast to CPU implementations, where reducing FLOPs often leads to higher performance, achieving near-peak performance on GPUs requires prioritizing the reduction of global-memory traffic.
Based on this understanding, we develop a two-stage algorithm,
in which contributions from Gaussian pairs at each grid point are first accumulated in registers or shared memory, followed by a single write of the aggregated result to global memory.
With such a strategy, for example, the global memory write associated with the density construction [Eq.~\eqref{eq:rho}] is reduced to its theoretical minimum, 
$N_\text{grid}$.
This also naturally introduces an additional level of parallelism over the grid points.
In the following, we provide more details of this GPU implementation of the multigrid FFTDF approach.

First, to exploit grid-level parallelism, the uniform grid is logically partitioned into 64-point ($4\times 4 \times 4$) grid blocks,
each mapped to a CUDA thread block (which contains 64 threads) across all our custom CUDA kernels.
For each grid block, we record the contributing Gaussian pairs,
i.e., those with nonzero overlap on the corresponding grid points.
The screening procedure is technically involved and is described in the Appendix.
The computational cost of this step is approximately 1.5 times that of a single SCF cycle, but it is performed only once for the entire energy and nuclear gradient evaluation.

\begin{algorithm}

	\var{\_\_global\_\_} $\rho(\mathbf{r})$;

	\In{$B_n$} {
	\var{\_\_shared\_\_} $\rho_\text{shared}(\mathbf{r}\in B_n)$;

	\var{n\_batch} \var{=} $\text{size}[\{\Lambda_p\}_{B_n}]$ \var{//} 64 \var{+} 1;

	\var{thread\_id} \var{=} \var{threadIdx.x};

	\For{\textup{\var{batch\_idx}} \KwTo \Range{\textup{\var{n\_batch}}}}{
	$p$ \var{=} \var{batch\_idx} \var{*} 64 \var{+} \var{thread\_id};

	Read $D_{\mu\nu \in \Lambda_p}$;

	\For{$\mathbf{r}$ \KwTo $B_n$}{
	\var{contracted} \var{=}  $\sum_{\mu\nu \in \Lambda_p}D_{\mu\nu} \Lambda_{p}(\mathbf{r})$;

	$\rho_\text{shared}(\mathbf{r})$ \var{+=} \var{reduce(contracted)};
	}
	}

	$\rho(\mathbf{r})$ \var{=} $\rho_\text{shared}(\mathbf{r})$
	}
	\caption{Workflow of electron density build}
	\label{alg:density}
\end{algorithm}

Next, we employ a two-stage parallelism for constructing the electron density [Eq.~\eqref{eq:rho}].
The procedure is outlined in Algorithm \ref{alg:density}. 
Within each CUDA thread block $B_n$ (assigned with a grid block),
we loop over the corresponding contributing set of primitive Gaussian shell pairs
$\{\Lambda_p\equiv g_{ab}^{l^a l^b}\}_{B_n}$ in batches.
In each batch, the 64 threads each process one shell pair and computes its contribution to the density in parallel.
Specifically, each thread loads into registers the subblock of the density matrix associated with the shell pair and contracts it with the result of Eq.~\eqref{eq:gab}, which is evaluated on the grid points (belonging to $B_n$) using the recursion relations Eqs.~\eqref{eq:exp_rr} and \eqref{eq:exp_rr1}.
When computing the kinetic energy density for meta-GGA functionals, $\Lambda_p(\mathbf{r})$ is replaced by $\bm{\nabla} g_a(\mathbf{r}) \cdot \bm{\nabla} g_b(\mathbf{r})$.
Unlike the CPU implementation, the polynomial prefactor in Eq.\eqref{eq:gab} is evaluated directly, without applying the binomial expansion in Eq.\eqref{eq:binomial}.
This is because forming the intermediate tensor
$\mathcal{D}$ in Eq.~\eqref{eq:rho1} would exceed the available GPU register capacity, 
causing register spilling and increased memory traffic, thereby significantly degrading the performance.
A reduction over the threads within the block $B_n$ is then performed using the CUB\cite{cccl} library to accumulate the partial density, which is stored in shared memory (denoted as $\rho_\text{shared}$).
Finally, $\rho_{\text{shared}}$ is written to the global density output $\rho$ in parallel, with each thread responsible for one grid point.
It is clear that the number of global memory writes is strictly at its theoretical minimum, $N_\text{grid}$, and that the global memory latency does
not affect the compute-dominant steps.
Together, these two factors are key to the high performance of our GPU implementation.

\begin{algorithm}

	\var{\_\_global\_\_} $\mathbf{F^\text{2e}}$;

	\In{$B_n$} {

	\var{\_\_shared\_\_} $v_\text{Hxc} (\mathbf{r} \in B_n)$;

	\var{n\_batch} \var{=} $\text{size}[\{\Lambda_p\}_{B_n}]$ \var{//} 64 \var{+} 1;

	\var{thread\_id} \var{=} \var{threadIdx.x};

	\For{\textup{\var{batch\_idx}} \KwTo \Range{\textup{\var{n\_batch}}}}{
		$p$ \var{=} \var{batch\_idx} \var{*} 64 \var{+} \var{thread\_id};

		$F_{\mu\nu\in \Lambda_p}^{\text{local}}$ \var{=} 0;

		\For{$\mathbf{r}$ \KwTo $B_n$}{
			$F_{\mu\nu\in \Lambda_p}^{\text{local}}$ \var{+=} $v_\text{Hxc} (\mathbf{r}) \Lambda_p (\mathbf{r})$;
		}
		\var{atomicAdd}($F^\text{2e}_{\mu\nu\in\Lambda_p}$, $F_{\mu\nu\in \Lambda_p}^{\text{local}}$);
	}
	}
	\caption{Workflow of two-electron Fock matrix build}
	\label{alg:xc}
\end{algorithm}

The Fock matrix build [Eq.~\eqref{eq:J}] follows a simpler workflow, as outlined in Algorithm~\ref{alg:xc}.
The real-space Hxc potential ($v_\text{Hxc}$) is first loaded into shared memory, enabling low-latency access by all threads within a thread block.
The primitive Gaussian shell pairs contributing to the current grid block are then processed in parallel. Each thread handles one shell pair, evaluates its values on the grid points using the recursion relations [Eqs.~\eqref{eq:exp_rr} and \eqref{eq:exp_rr1}], and contracts the result with $v_\text{Hxc}$ to form the local Fock matrix contribution $F^\text{local}$.
For meta-GGA functionals, we also add contributions from the kinetic energy density, 
\begin{equation}
F_{\mu\nu\in \Lambda_p}^{\text{local}} \,\texttt{+=}\, \frac12 v_\text{xc}^\text{mGGA}(\mathbf{r}) \bm{\nabla} g_a(\mathbf{r}) \cdot \bm{\nabla} g_b(\mathbf{r}) \;. \nonumber
\end{equation}
Finally, $F^\text{local}$ is reduced to the global Fock matrix (in the primitive basis; the transformation from the primitive to the contracted basis is performed outside the kernel via efficient sparse matrix multiplications) using \texttt{atomicAdd}.
Two aspects of this design are critical for performance. First, all threads in a block simultaneously read $v_\text{Hxc}$ at the same grid point from shared memory, which effectively minimizes shared-memory access latency compared to unordered access.
Second, each thread updates distinct shell-pair elements in the global Fock matrix, thereby minimizing write contention and preventing the severe latency penalties that would otherwise arise from contended \texttt{atomicAdd} operations.

\begin{algorithm}

	\var{\_\_global\_\_} $\bm{\nabla}_A E$; \Comment{$A$ iterates over atoms}

	\In{$B_n$} {
	\var{\_\_shared\_\_} $v_\text{Hxc} (\mathbf{r} \in B_n)$;

	\var{n\_batch} \var{=} $\text{size}[\{\Lambda_p\}_{B_n}]$ \var{//} 64 \var{+} 1;

	\var{thread\_id} \var{=} \var{threadIdx.x};

	\For{\textup{\var{batch\_idx}} \KwTo \Range{\textup{\var{n\_batch}}}}{
	$p$ \var{=} \var{batch\_idx} \var{*} 64 \var{+} \var{thread\_id};

	Read $D_{\mu\nu \in \Lambda_p}$;

	[$\bm{\nabla}_A E^\text{local}, \bm{\nabla}_B E^\text{local}$] \var{=} 0;

	\For{$\mathbf{r}$ \KwTo $B_n$}{
	$\bm{\nabla}_A E^\text{local}$ \var{+=}  $\sum_{\mu\nu \in \Lambda_p}D_{\mu\nu} v_\text{Hxc}(\mathbf{r}) \bm{\nabla}_A \Lambda_p(\mathbf{r})$;

	$\bm{\nabla}_B E^\text{local}$ \var{+=}  $\sum_{\mu\nu \in \Lambda_p}D_{\mu\nu} v_\text{Hxc}(\mathbf{r}) \bm{\nabla}_B \Lambda_p(\mathbf{r})$;
	}
	\AtomicAdd{$\bm{\nabla}_A E$, $\bm{\nabla}_A E^\textup{local}$};

	\AtomicAdd{$\bm{\nabla}_B E$, $\bm{\nabla}_B E^\textup{local}$};

	}
	}
	\caption{Workflow of energy gradient}
	\label{alg:gradient}
\end{algorithm}

Finally, for the computation of nuclear gradients, the contribution from the Hxc potential is evaluated in a way analogous to Algorithm~\ref{alg:xc}, except that gradients of Gaussian pairs with respect to nuclear coordinates are evaluated (denoted as $\bm{\nabla}_A \Lambda_p$), and that the energy gradient is obtained directly by contracting the density matrix,
$v_\text{Hxc}$, and $\bm{\nabla}_A \Lambda_p$, as shown in Algorithm~\ref{alg:gradient} (lines 10–12). 
For meta-GGA functionals, the contribution from the kinetic energy density is computed as
\begin{equation}
\bm{\nabla}_A E^\text{local} \; \text{\texttt{+=}} \sum_{\mu\nu \in \Lambda_p}D_{\mu\nu} v_\text{xc}^\text{mGGA}(\mathbf{r})\bm{\nabla}_A \bigg(\bm{\nabla} g_a(\mathbf{r}) \cdot \bm{\nabla} g_b(\mathbf{r})\bigg)\;, \nonumber
\end{equation}
which involves second-order derivatives of the basis functions.

\section{Results and discussions \label{sec:result}}

\begin{table*}
	\caption{
		Wall times (in seconds) for single SCF iterations on average
		and nuclear gradient calculations within restricted KS-DFT for various systems, using the
		PBE functional and the GTH-PADE pseudopotentials.
	}\label{tab:timing}
	\scriptsize
	\centering
	\begin{tabular*}{1.0\textwidth}{@{\extracolsep{\fill}}lrcccccc}
		\hline\hline
		&&\multicolumn{3}{c}{1 SCF iteration on average} & \multicolumn{3}{c}{Nuclear gradient}  \\
		\cline{3-5}\cline{6-8}
		\multirow{2}{*}{System} & \multirow{2}{*}{$N_\text{basis}$}  &
		\textsc{GPU4PySCF} & \textsc{GPU4PySCF} & \textsc{PySCF} &
		\textsc{GPU4PySCF} & \textsc{GPU4PySCF} & \textsc{PySCF}\\
		& & H100 & A100 & 28-core CPU & H100 & A100 & 28-core CPU \\\hline
		\multicolumn{8}{c}{GTH-TZV2P basis set} \\\hline
		32\ce{H2O} &   1280 &   0.06 &   0.08 &   0.52 &   0.31 &   0.39 &   1.25 \\
		64\ce{H2O} &   2560 &   0.13 &   0.21 &   1.83 &   0.47 &   0.70 &   3.33 \\
		128\ce{H2O} &   5120 &  0.43 &   0.79 &   8.97 &   1.15 &   1.99 &   9.03 \\
		256\ce{H2O} &  10240 &  1.86 &   3.80 &  52.52 &   4.02 &   7.60 &  29.37 \\
		512\ce{H2O} &  20480 & 13.42 &  30.58 & 226.28 &  17.07 &  38.11 & 108.30 \\
		\\
		1 $\times$ 1 $\times$ 1 benzene &    186 &   0.32 &   0.08 &   0.10 &   0.53 &   0.24 &   0.16 \\
		2 $\times$ 2 $\times$ 2 benzene &   1488 &   0.37 &   0.15 &   0.95 &   0.70 &   0.58 &   1.35 \\
		3 $\times$ 3 $\times$ 3 benzene &   5022 &   0.94 &   0.91 &   7.92 &   1.83 &   2.73 &   8.54 \\
		4 $\times$ 4 $\times$ 4 benzene &  11904 &   3.44 &   6.38 &  56.95 &   7.29 &  13.04 &  32.68 \\
		\hline
		\multicolumn{8}{c}{GTH-DZVP basis set} \\\hline
		2 $\times$ 2 $\times$ 2 diamond &    416 &   0.07 &   0.09 &   0.23 &   0.51 &   0.54 &   1.48 \\
		3 $\times$ 3 $\times$ 3 diamond &   1404 &   0.12 &   0.22 &   1.00 &   0.61 &   0.93 &   5.56 \\
		4 $\times$ 4 $\times$ 4 diamond &   3328 &   0.31 &   0.62 &   4.13 &   2.19 &   3.39 &  15.95 \\
		5 $\times$ 5 $\times$ 5 diamond &   6500 &   0.98 &   1.91 &  13.18 &   7.37 &  11.45 &  38.19 \\
		6 $\times$ 6 $\times$ 6 diamond &  11232 &   2.79 &   5.81 &  46.62 &  18.56 &  28.36 &  84.56 \\
		\\
		2 $\times$ 2 $\times$ 2 LiF &    864 &   0.64 &   0.25 &  0.65 &  1.11 &  1.19 &  11.37 \\
		3 $\times$ 3 $\times$ 3 LiF &   2916 &   0.53 &   0.63 &  4.01 &  1.97 &  2.37 &  41.76 \\
		4 $\times$ 4 $\times$ 4 LiF &   6912 &   2.04 &   2.14 & 24.79 &  6.16 &  9.14 & 119.94 \\
		\hline
	\end{tabular*}
\end{table*}

\begin{figure*}
	\includegraphics[width=\textwidth]{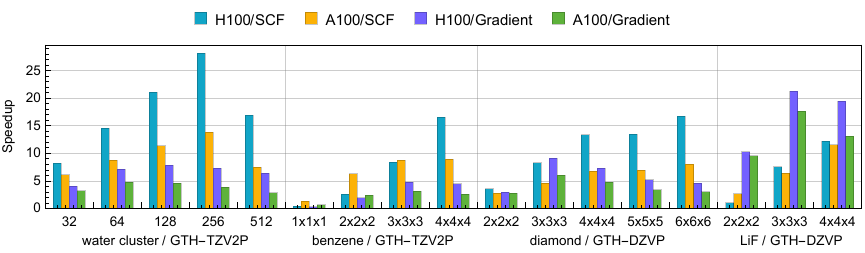}
	\caption{Speedups of \textsc{GPU4PySCF} on NVIDIA H100 and A100 GPUs for a single SCF iteration and nuclear gradient calculation, relative to the corresponding \textsc{PySCF} CPU timings reported in Table \ref{tab:timing}.}
	\label{fig:scf_grad_speedup}
\end{figure*}

\begin{table*}
	\caption{
		Wall times (in seconds) for one Fock build on average using different packages.
		Single-core CPU times are reported in parentheses for a subset of water clusters. 
	}\label{tab:timing_fock}
	\scriptsize
	\centering
	\begin{tabular*}{1.0\textwidth}{@{\extracolsep{\fill}}lrccccc}
		\hline\hline
		\multirow{2}{*}{System} & \multirow{2}{*}{$N_\text{basis}$} &
		\textsc{GPU4PySCF} & \textsc{GPU4PySCF} & CP2K & \textsc{PySCF} & CP2K \\
		& & H100 & A100 & A100 & 28-core CPU & 28-core CPU\\\hline

		\multicolumn{7}{c}{GTH-TZV2P basis set} \\\hline
		32\ce{H2O} &   1280 &    0.04 &   0.05 &   0.32 &   0.32 &   0.21 \\
		64\ce{H2O} &   2560 &    0.06 &   0.10 &   0.47 &   0.68 (12.95) &   0.42 (9.12) \\
		128\ce{H2O} &   5120 &   0.12 &   0.23 &   0.81 &   1.57 (24.69) &   0.85 (20.22) \\
		256\ce{H2O} &  10240 &   0.33 &   0.60 &   1.56 &   3.56 (41.01) &   1.73 (42.66) \\
		512\ce{H2O} &  20480 &   0.94 &   5.48 &   3.32 &   8.08 (87.00) &   5.54 (68.24) \\
		1024\ce{H2O} & 40960 & --\footnote{Computation not completed due to insufficient GPU memory.\label{note:oom}} & --\footref{note:oom} & --\footref{note:oom} & 23.50 & 10.36\\
		2048\ce{H2O} & 81920 & --\footref{note:oom} & --\footref{note:oom} & --\footref{note:oom} & 68.49 &  16.78\\
		\\
		1 $\times$ 1 $\times$ 1 benzene &    186 &   0.31 &   0.07 &   0.18 &   0.09 &   0.08 \\
		2 $\times$ 2 $\times$ 2 benzene &   1488 &   0.33 &   0.11 &   0.54 &   0.71 &   0.54 \\
		3 $\times$ 3 $\times$ 3 benzene &   5022 &   0.64 &   0.36 &   1.91 &   2.74 &   1.81 \\
		4 $\times$ 4 $\times$ 4 benzene &  11904 &   1.17 &   1.47 &   4.72 &   7.08 &   4.25 \\
		\hline
		\multicolumn{7}{c}{GTH-DZVP basis set} \\\hline
		2 $\times$ 2 $\times$ 2 diamond &    416 &   0.05 &   0.08 &   0.14 &   0.21 &   0.19 \\
		3 $\times$ 3 $\times$ 3 diamond &   1404 &   0.09 &   0.18 &   0.25 &   0.80 &   0.40 \\
		4 $\times$ 4 $\times$ 4 diamond &   3328 &   0.19 &   0.41 &   0.52 &   2.05 &   0.98 \\
		5 $\times$ 5 $\times$ 5 diamond &   6500 &   0.41 &   0.86 &   0.99 &   4.36 &   1.88 \\
		6 $\times$ 6 $\times$ 6 diamond &  11232 &   0.77 &   1.58 &   1.96 &   8.25 &   3.38 \\
		\\
		2 $\times$ 2 $\times$ 2 LiF &    864 &   0.63 &   0.23 &   0.34 &   0.54 &   0.73 \\
		3 $\times$ 3 $\times$ 3 LiF &   2916 &   0.46 &   0.49 &   0.96 &   2.15 &   2.59 \\
		4 $\times$ 4 $\times$ 4 LiF &   6912 &   1.47 &   1.07 &   2.55 &   6.87 &   6.68 \\
		\hline
	\end{tabular*}
\end{table*}

\begin{figure*}
	\includegraphics[width=\textwidth]{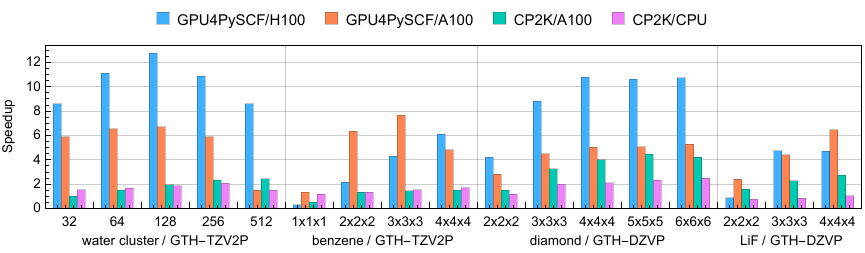}
	\caption{Speedups of \textsc{GPU4PySCF} on NVIDIA H100 and A100 GPUs, and of CP2K on A100 GPUs and CPUs, for a single Fock build, relative to the corresponding \textsc{PySCF} CPU timings reported in Table \ref{tab:timing_fock}.
	}
	\label{fig:fock_speedup}
\end{figure*}

\begin{figure}
	\begin{subfigure}{0.45\textwidth}
		\includegraphics[width=\textwidth]{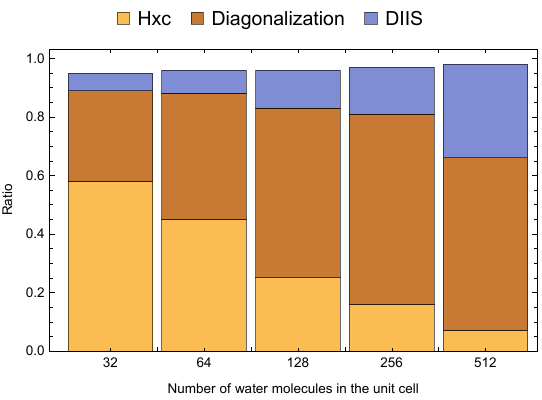}
		\caption{SCF}
		\label{fig:scf_ratio}
	\end{subfigure}
	\begin{subfigure}{0.45\textwidth}
		\includegraphics[width=\textwidth]{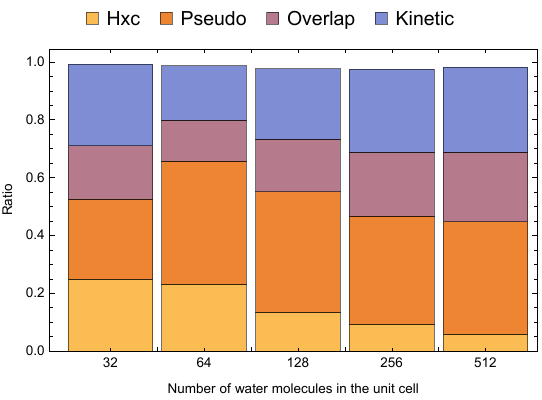}
		\caption{Gradient}
		\label{fig:grad_ratio}
	\end{subfigure}
	\caption{The computational time ratio for subroutines in an SCF cycle (a) and nuclear gradient calculation (b) for water clusters, benchmarked on H100 GPUs.
		In the legends, ``Hxc'' denotes Hxc potential,
		and ``Pseudo'' denotes pseudopotential.}
	\label{fig:ratio}
\end{figure}

\begin{figure}
	\begin{subfigure}{0.4\textwidth}
		\includegraphics[width=\textwidth]{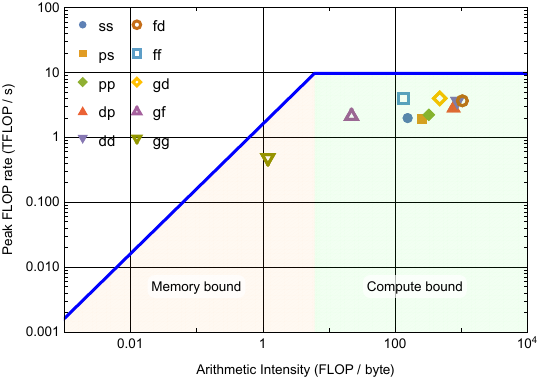}
		\caption{electron density build kernel}
		\label{fig:density}
	\end{subfigure}

	\begin{subfigure}{0.4\textwidth}
		\includegraphics[width=\textwidth]{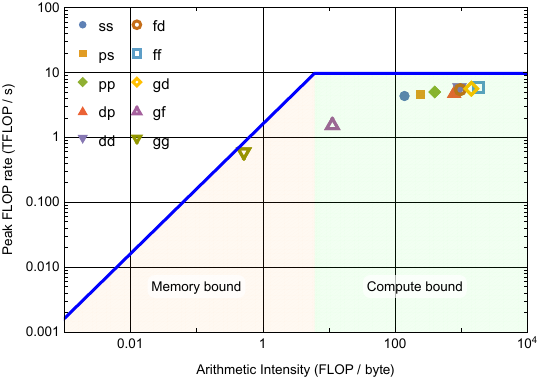}
		\caption{effective potential build kernel}
		\label{fig:xc}
	\end{subfigure}

	\begin{subfigure}{0.4\textwidth}
		\includegraphics[width=\textwidth]{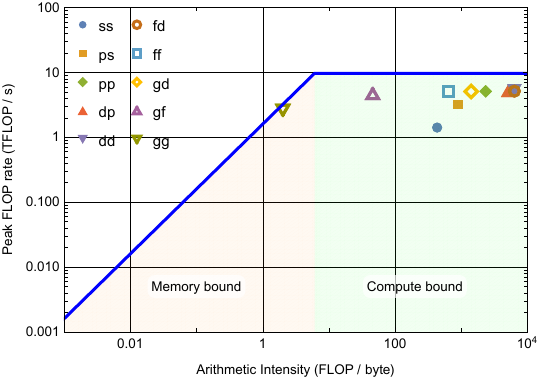}
		\caption{effective potential gradient kernel}
		\label{fig:grad}
	\end{subfigure}
	\caption{FLOP performance of the custom CUDA kernels analyzed using the roofline model benchmarked on the NVIDIA A100 GPU.
		The solid blue line represents the official peak FP64 FLOP rate of 9.7 TFLOP/s with no bandwidth constraint (horizontal)
		and the peak FP64 FLOP rate constrained by the peak memory bandwidth of 1.6 TB/s (diagonal).
		The theoretical arithmetic intensity of 6.1 FLOP/byte marks the boundary between the memory-bound zone and the compute-bound zone for the A100 GPU. The benchmark calculations were performed for a 32-water cluster at the PBE/GTH-cc-pVQZ level of theory.}

\end{figure}

We have implemented the multigrid FFTDF approach in \textsc{PySCF} and \textsc{GPU4PySCF} for CPUs and NVIDIA GPUs, respectively. 
All results were obtained using forks of \textsc{PySCF}\footnote{https://github.com/fishjojo/pyscf/tree/multigrid2\_large\_system} and \textsc{GPU4PySCF}\footnote{https://github.com/Walter-Feng/gpu4pyscf/tree/multi-grid/publish}.
In this section, we examine their performance.
The benchmark systems include water clusters, crystalline benzene, diamond, and LiF.
The geometries of the water clusters were taken from the CP2K benchmark suit,\cite{vandevondele2005quickstep}
while those of crystalline benzene and LiF were adopted from Ref.\citenum{wangFastScalableGPUAccelerated2024}.
Conventional diamond supercells were constructed using the Atomic Simulation Environment (ASE).\cite{ase-paper}
All calculations were performed within restricted KS-DFT using the PBE\cite{Perdew1996-hz} functional and the GTH-PADE\cite{hartwigsen1998relativistic} pseudopotentials.
For water and benzene, the GTH-TZV2P\cite{vandevondele2007gaussian} basis set was employed,
whereas the GTH-DZVP\cite{vandevondele2007gaussian} basis set was used for diamond and LiF.
A plane wave kinetic energy cutoff of 140 a.u. was used.
The GTO screening threshold $\tau$ in Eq.~\eqref{eq:cut} was set to $10^{-6}$,
which gives an accuracy comparable to that obtained by setting the keyword EPS\_DEFAULT to $10^{-12}$ in CP2K, the program we compare our code with.
Benchmark calculations were run on NVIDIA A100 (80 GB) and H100 (80 GB) GPUs, as well as on Intel Cascade Lake 8276 CPUs.

First, we report the wall times for SCF iterations and nuclear gradient calculations in Table \ref{tab:timing} and Figure \ref{fig:scf_grad_speedup}.
For SCF iterations, \textsc{GPU4PySCF} running on an NVIDIA A100 GPU achieves a speedup of $4$--$10\times$ relative to \textsc{PySCF} running on 28 CPU cores.
Using an H100 GPU provides an additional $2\times$ speedup over the A100, consistent with the ratio of their peak FP64 throughputs.

In Table \ref{tab:timing_fock} and Figure \ref{fig:fock_speedup}, we further compare the Fock build times of our implementations with those of CP2K, which is widely regarded as one of the most efficient codes for Gaussian-Plane-Wave KS-DFT.
For CP2K, the GPU benchmarks were performed using the Docker image version 2025.1, while the CPU benchmarks [run with MPI (Message Passing Interface)] employed the 2025.2 Docker image built for Intel Cascade Lake CPUs with AVX-512 support.
On CPUs, \textsc{PySCF} is slower than CP2K by a factor of approximately two.
This can be attributed to less efficient parallelization, as reflected by the comparable single-core CPU timings reported in parentheses in Table \ref{tab:timing_fock}.
In contrast, \textsc{GPU4PySCF}
exhibits a significant performance advantage for water and benzene,
giving a roughly $3\times$ speedup over the GPU CP2K implementation on A100 GPUs.
This advantage is smaller for diamond and LiF.
This may be due to the larger fraction of diffuse Gaussian functions in these systems, which leads to a larger amount of redundant computation of Gaussian pair prefactors across different grid blocks.

For nuclear gradient calculations, the GPU speedup is generally smaller than that observed for SCF iterations because the calculation of one-electron integrals has not been fully optimized for GPUs.
For example, in water clusters studied here that have more than 1000 basis functions, the Hxc potential accounts for less than 25\%  of the total gradient wall time, 
as shown in Fig.~\ref{fig:grad_ratio}.
A similar trend is observed in the energy evaluations, because as the system size increases, the eigensolver and DIIS\cite{pulayConvergenceAccelerationIterative1980} steps dominate the computational cost (see Fig.~\ref{fig:ratio}a).
We leave further optimization of these remaining components beyond the Fock build to future work.
Nevertheless, even in its current form, our GPU-accelerated KS-DFT implementation is highly efficient for medium-sized systems. For example, the energy and nuclear gradient calculation for a 256-molecule water cluster (with 10,000 basis functions) finishes in only 30 seconds on an H100 GPU.

We next present the roofline analysis for our GPU-accelerated multigrid FFTDF implementation.
For the electron density build, as shown in Fig. \ref{fig:density},
the kernels for computing $s$-shell Gaussian pair contributions [denoted as $(s|s)$]
achieve approximately 50\% of the peak FP64 throughput on A100 GPUs, while the $(d|d)$ and $(f|f)$ kernels reach about 70\% of peak performance. The higher effective FLOP rate for larger angular momenta arises from the additional arithmetic associated with generating more Cartesian components within a shell of Gaussian orbitals, whose instructions can effectively hide shared-memory latency during the reduction. This improves utilization of the FP64 cores within each streaming multiprocessor (SM).
In contrast, a performance degradation is observed when $g$-shell orbitals are involved. In this case, the increased number of intermediate variables exceeds the available register capacity, requiring the use of global memory and rendering the kernels memory-bandwidth-bound. The primary bottleneck is the reduction of the partial densities in shared memory at each recursion step (Algorithm~\ref{alg:density}, line 11). Because this reduction requires thread synchronization, the latency associated with shared-memory access can no longer be hidden by instruction-level parallelism.

The kernels for constructing the Coulomb matrix are more compute-bound, as shown in Fig.~\ref{fig:xc}.
This is because the Hxc potential is stored in shared memory and can be broadcast to all threads during each recursion step (Algorithm~\ref{alg:xc}, line 10) without requiring synchronization,
resulting in minimal shared memory traffic.
Consequently, these kernels can achieve approximately 80\% of the peak FP64 throughput on A100 GPUs.
A similar performance degradation is again observed when $g$-shell orbitals are involved, reflecting the increased register pressure associated with higher angular momentum.

The kernels for computing the nuclear gradient exhibit even higher arithmetic intensity (see Fig.~\ref{fig:grad}),
as the density matrix is contracted with the Fock matrix on the fly,
although we observe that the low angular momentum kernels,
such as the $(s|s)$ and $(p|s)$ kernels, do not fully saturate the compute units.

\section{Conclusion \label{sec:conclusion}}

We have described our implementation of the multigrid Gaussian-Plane-Wave (FFTDF) Kohn-Sham DFT algorithm that enables highly-performant KS-DFT calculations on both CPUs and GPUs. Currently our CPU implementation supports KS-DFT simulations with up to 100,000 basis functions on a single shared memory node, while the GPU implementation provides an order-of-magnitude speedup over the single-node CPU performance, subject to available device memory. Notably, by implementing two-level parallelism, our GPU kernels achieve approximately 80\% of the peak FP64 throughput, leading to state-of-the-art performance without degradation for orbitals up to the $f$ shell.

The implementation in \pyscf establishes a strong open-source foundation for many applications in computational chemistry and computational materials science, including large-scale materials screening and {\em ab initio} molecular dynamics. The full potential of our approach can be further realized through \pyscf's support for QM/MM and quantum embedding methods, enabling significantly larger quantum mechanical regions to be treated at the DFT level. Our GPU implementation strategy may also be useful in the development of fast Gaussian-Plane-Wave exact exchange algorithms. We will report on these developments in future work.

\section*{Acknowledgements}

RL (development of GPU multigrid code) and GKC (project supervision) were supported by the US Department of Energy, Office of Science, through Award No. DE-SC0023318.
XZ (development of CPU multigrid code) was supported by the Center for Molecular Magnetic Quantum Materials, an Energy Frontier Research Center funded by the U.S. Department of Energy, Office of Science, Basic Energy Sciences under Award No. DE-SC0019330.
Software packaging, integration, and infrastructure development in \pyscf was supported by the US National Science Foundation under Award 2513474.

This research used resources of the National Energy Research Scientific Computing Center (NERSC), a U.S. Department of Energy Office of Science User Facility located at Lawrence Berkeley National Laboratory, operated under Contract No. DE-AC02-05CH11231 using NERSC award ERCAP-0024087.
Part of the computations were conducted in the Resnick High Performance Computing Center, a facility supported by Resnick Sustainability Institute at the California Institute of Technology.
GKC is a Simons Investigator in Physics.

\appendix*
\section*{Appendix}

\subsubsection{GTO screening on GPUs}

The screening procedure consists of two steps:
identifying the contributing Gaussian pairs, and distributing these pairs to the grid blocks with which they overlap.
The corresponding pseudocode is given in Algorithms~\ref{alg:screening} and \ref{alg:mapping}, respectively.
In the former step, primitive Gaussian pairs are processed in parallel and their spatial boundaries are computed.
Pairs that overlap with the reference unit cell have their indices and boundaries recorded for subsequent grid block assignment.
In the latter step, we first count the number of contributing pairs for each grid block (\texttt{n\_pairs\_per\_block}) based on the pairs' boundaries.
The indices of the contributing pairs are then populated into a list (\texttt{indices\_on\_blocks}).
Together, they identify the Gaussian pairs contributing to each grid block.
To maximize compute throughput, the pair indices are first pooled in shared memory,
and then written to global memory only when the buffer is full (Algorithm~\ref{alg:mapping}, lines 22--29). To achieve a more fine-grained screening,
the screened Gaussian pairs are evaluated on the corresponding grid blocks, and are removed from the list if the absolute values do not exceed the accuracy threshold $\tau$ (Algorithm~\ref{alg:mapping}, line 30-41).

\begin{algorithm}
	Count number of contributing pairs \var{$N_\text{pair}$};

	\var{\_\_global\_\_} \var{pair\_index[$N_\text{pair}$]};

	\var{\_\_global\_\_} \var{boundary[$N_\text{pair}$]};

	\var{\_\_global\_\_} \var{recorded\_count};

	\In{$B_x, B_y$} {

		\var{$(\mu, \mathbf{T}_1)$} \var{=} \var{threadIdx.x} \var{+} $B_x$ \var{*} \var{blockDim.x};

		\var{$(\nu, \mathbf{T}_2)$} \var{=} \var{threadIdx.y} \var{+} $B_y$ \var{*} \var{blockDim.y};

		\var{\_\_shared\_\_} \var{offset};

		\var{is\_valid} \var{=} 0;

		Compute boundaries $\mathbf{q}_\text{min}$, $\mathbf{q}_\text{max}$ in fractional coordinates);

		\If{$q_{\textup{min},\tau} < 1$ \textup{and} $q_{\textup{max},\tau} > 0$ \textup{for all} $\tau \in \{x,y,z\}$}{
			\var{is\_valid} \var{=} 1;
		}

        Change $\mathbf{q}_\text{min}$, $\mathbf{q}_\text{max}$ to grid point indices, and extend to block boundaries (multiple of 4 in grid point indices)

		\var{prefix\_sum} \var{=} 0;

		\var{aggregated} \var{=} 0;

		\var{ExclusiveSum(is\_valid, prefix\_sum, aggregated)};

		\If{\textup{master thread}}{
			\var{offset} = \var{atomicAdd}(\var{recorded\_count}, \var{aggregated});
		}

		\var{index} = \var{offset} + \var{prefix\_sum};

		\var{pair\_index[index]} = $(\mu, \nu, \mathbf{T}_1, \mathbf{T}_2)$;

		\var{boundary[index]} = $(\mathbf{q}_\text{min}, \mathbf{q}_\text{max})$;

	}
	\caption{Workflow of Gaussian pair screening}
	\label{alg:screening}
\end{algorithm}

\begin{algorithm}
	\var{\_\_global\_\_} \var{n\_pairs\_per\_block[n\_blocks]};

	\In{$B_n$} {

		\var{count = 0};

		\For{\textup{\var{batch}} \KwTo $\{\Lambda_p\}$}{

			\If{$B_n \in (\mathbf{q}_\text{min}, \mathbf{q}_\text{max})$}{

				\var{count++};

			}
		}

		\var{n\_pairs\_per\_block[$B_n$]} = \var{reduce}(\var{count});
	}

	\var{\_\_global\_\_} \var{offsets = \{0\} + cumsum(n\_pairs\_per\_block)};

	\var{\_\_global\_\_} \var{indices\_on\_blocks[offsets[-1]]};

	\In{$B_n$} {

		\var{\_\_shared\_\_} \var{pool[pool\_size]};

		\var{n\_pooled = 0};

		\var{offset = offsets[$B_n$]};

		\For{\textup{\var{batch}} \KwTo $\{\Lambda_p\}$}{

			\var{p = batch * batch\_size + threadIdx.x};

			\var{is\_valid = 0};

			\If{$B_n \in (\mathbf{q}_{min}, \mathbf{q}_{max})$}{
				\var{is\_valid = 1};
			}

			\var{aggregated = 0};

			\var{prefix\_sum = 0};

			\var{ExclusiveSum(is\_valid, prefix\_sum, aggregated)};

			\If{\textup{\var{aggregated}} $+$ \textup{\var{n\_pooled}} $>$ \textup{\var{pool\_size}}}{

				Store pooled pairs to \var{indices\_on\_blocks};

				\var{offset += aggregated};

				\var{n\_pooled = 0};
			}

			\var{pool[n\_pooled + prefix\_sum] = p};

			\var{n\_pooled += aggregated};

		}

		\If{\textup{\var{n\_pooled}} $> 0$}{

			Store pooled pairs to \var{indices\_on\_blocks};

		}

	}
    \In{$B_n$} {

        \var{n\_batch} \var{=} $\text{size}[\{\Lambda_p\}_{B_n}]$ \var{//} 64 \var{+} 1;
        
        \var{offset = offsets[$B_n$]};

        \For{\textup{\var{batch\_idx}} \KwTo \Range{\textup{\var{n\_batch}}}}{
            $p$ \var{=} \var{batch\_idx} \var{*} 64 \var{+} \var{thread\_id};
        
               \var{max\_value = 0};
        
            \For{$\mathbf{r}$ \KwTo $B_n$}{
                   \var{max\_value} \var{=} \var{max}(\var{max\_value}, $|4 \pi r^2 \Lambda_p (\mathbf{r})|$);
            }
            \If{\textup{\var{max\_value} \var{<} $\tau$}}{
                \var{indices\_on\_blocks}\var{[offset + $p$]} \var{= -1};
        
                \var{atomicAdd}(\var{n\_pairs\_per\_blocks}\var{[$B_n$]}, \var{-1});
            }
        }

	}
    Filter negative pair indices in \var{indices\_on\_blocks};
	\caption{Workflow of mapping Gaussian pair to grid blocks}
	\label{alg:mapping}
\end{algorithm}

\subsubsection{Non-orthogonal lattices}
We denote the three lattice vectors as $\{\mathbf{R}_1, \mathbf{R}_2, \mathbf{R}_3\}$, and the corresponding fractional coordinates of the grid points are noted as $\{u_i, v_j, w_k\}$. Given a starting point with fractional coordinate $\{u_0, v_0, w_0\}$ and absolute coordinate $\mathbf{r}_0 = u_0\mathbf{R}_1 + v_0\mathbf{R}_2 + w_0\mathbf{R}_3$, the fractional coordinate of each point can be expressed simply as $u_i = i /N_1 + u_0$, $v_j = j / N_2 + v_0$, $w_k = k / N_3 + w_0$, where $N_1$, $N_2$, $N_3$ are the number of grid points along each lattice vector. The recursion form for the computation of a Gaussian pair becomes
\begin{align}
&\begin{aligned} 
	g_{ab}^{00}(u_{i+1}, 0, 0) & = g_{ab}^{00}(u_i, 0,0) e^{-(2i+1)\alpha_P |d\mathbf{R}_1|^2} e^{-2\alpha_P (\mathbf{r}_0 - \mathbf{P}) \cdot d\mathbf{R}_1}
\end{aligned}\label{eq:exp_rr1_non_orthogonal_r1}\\
&\begin{aligned} 
	g_{ab}^{00}(u_i, v_{j+1}, 0) & = g_{ab}^{00}(u_i, v_j, 0) e^{-(2j+1)\alpha_P |d\mathbf{R}_2|^2} e^{-2\alpha_P (\mathbf{r}_0 - \mathbf{P}) \cdot d\mathbf{R}_2} \\
    & \quad \quad e^{-2 u_i \alpha_p d\mathbf{R}_1 \cdot d\mathbf{R}_2}\\ 
\end{aligned}\label{eq:exp_rr1_non_orthogonal_r2}\\
&\begin{aligned} 
	g_{ab}^{00}(u_i, v_j, w_{k+1}) & = g_{ab}^{00}(u_i, v_j, w_k) e^{-(2k+1)\alpha_P |d\mathbf{R}_3|^2} \\
    & \qquad e^{-2\alpha_P (\mathbf{r}_0 - \mathbf{P}) \cdot d\mathbf{R}_3} \\
    & \qquad e^{-2 u_i \alpha_p d\mathbf{R}_1 \cdot d\mathbf{R}_3}e^{-2 v_j \alpha_p d\mathbf{R_2} \cdot d\mathbf{R}_3}\; \\ 
\end{aligned}\label{eq:exp_rr1_non_orthogonal_r3},
\end{align}
with initial condition
\begin{equation}
g_{ab}^{00}(0, 0, 0) = e^{-\alpha_P |\mathbf{r}_0 - \mathbf{P}|^2}.
\end{equation}

\section*{References}
\bibliography{references}

\end{document}